\documentclass[aps,prd,showkeys,nofootinbib,superscriptaddress,tightenlines,11pt]{revtex4-2}

\usepackage[a-2u]{pdfx}
\usepackage[utf8]{inputenc}
\usepackage{lmodern}

\usepackage{amsmath}        % extensions for typesetting of math
\usepackage{amsfonts}       % math fonts
\usepackage{dsfont}			% math fonts like R, C, N etc,
\usepackage{amsthm}         % theorems, definitions, etc.
\usepackage{bbding}         % various symbols (squares, asterisks, scissors, ...)
\usepackage{bm}             % boldface symbols (\bm)
\usepackage{graphicx}       % embedding of pictures
\usepackage{fancyvrb}       % improved verbatim environment
\usepackage{natbib}         % citation style AUTHOR (YEAR), or AUTHOR [NUMBER]
\setcitestyle{square,citesep={;}}
\usepackage{dcolumn}        % improved alignment of table columns
\usepackage{booktabs}       % improved horizontal lines in tables
\usepackage{paralist}       % improved enumerate and itemize
\usepackage{xcolor}         % typesetting in color
\usepackage{physics}		% physical expressions
\usepackage{booktabs}
\usepackage{multirow}
\usepackage{xcolor}
\usepackage{hyphenat}
\usepackage{float}        % Exact placing of tables
\usepackage{tikz}
\usetikzlibrary{intersections}
\usepackage{pgfplots}
\usepgfplotslibrary{fillbetween}
\usepackage{subcaption}
\hypersetup{unicode}
\hypersetup{breaklinks=true}

\pgfplotsset{compat=1.18}

\theoremstyle{plain}

\DeclareMathOperator{\sgn}{sgn}

\newcommand{\goto}{\rightarrow}

\newcommand{\weyl}{\xrightarrow{\text{Weyl}}}

\newcommand{\ud}{\mathrm{d}}
\newcommand{\integral}[1]{\ensuremath\int\ud^{#1}x\,}

\newcommand{\ucharles}{Faculty of Mathematics and Physics, Charles University, V Hole\v{s}ovi\v{c}k\'{a}ch 2, 18000 Prague 8, Czech Republic}

\begin{document}

\title{Classical gravitational anomalies of Liouville theory}

\author{Pavel Haman}
\email{pavel.haman@mff.cuni.cz}
\affiliation{\ucharles}
\author{Alfredo Iorio}
\email{alfredo.iorio@mff.cuni.cz}
\affiliation{\ucharles}

\date{\today}

\begin{abstract}
We show that for classical Liouville field theory, diffeomorphism invariance, Weyl invariance and locality cannot hold together. This is due to a genuine Virasoro center, present in the theory, that leads to an energy\hyp{}momentum tensor with non-tensorial conformal transformations, in flat space, and with a non-vanishing trace, in curved space. Our focus is on a field-independent term, proportional to the square of the Weyl gauge field, $W_\mu W^\mu$, that makes the action Weyl-invariant and was disregarded in previous investigations of Weyl and conformal symmetry. We show this term to be related to the classical center of the Virasoro algebra. The mechanism uncovered here is a classical version of the quantum anomalous phenomenon: the generalization to curved space only allows to keep one of the two symmetries enjoyed by the flat space theory, either Lorentz (diffeomorphism) or conformal invariance.
\end{abstract}


\maketitle

%%%%%%%%%%%%%%%%%%%%%%%%%%%%%%%%%%%%%%%%%%%%%%%%%%%%%%%%%%%%%%%%%%%%%%%%
% ``For Liouville classical theory diffeomorphic invariance, Weyl invariance and locality cannot hold together'', a key role of the Virasoro classical center should be visible.
%%%%%%%%%%%%%%%%%%%%%%%%%%%%%%%%%%%%%%%%%%%%%%%%%%%%%%%%%%%%%%%%%%%%%%%%

Liouville field theory, with flat space action\footnote{Here we use $\eta_{\mu \nu} = {\rm diag} (+1,-1)$.}
\begin{equation} \label{eq:FlatLiouvilleTheory}
  A_{\scriptscriptstyle{L}}[\Phi] = \int \dd[2]{x}
  \qty(
    \frac{1}{2}\partial_\mu\Phi\partial^\mu\Phi
    - \frac{m^2}{\beta^2}e^{\beta\Phi}) \,,
\end{equation}
is an exactly solvable two-dimensional model that enjoys a prominent role in many fields of the theoretical and mathematical investigations. Among those, the geometry of surfaces \citep{Liouville_1853}, two-dimensional (quantum) gravity, see, e.g., \citep{QuantumGravity_Christensen} and \citep{DanielRuzziconi2022}, string theory, see, e.g.,  \citep{Polyakov_1981}, conformal field theories, such as the Wess-Zumino-Witten and the Toda models, see e.g., \citep{CFTDiFrancesco}, and therefore the AdS/CFT correspondence \citep{TasiOnAdsCft2016}. It is then of great importance to know its symmetries in all details, already at the classical level.

In particular, Liouville theory, besides being Lorentz invariant, is known to enjoy full (global) conformal symmetries in flat space, hence it belongs to the cases studied in \citep{WeylGauging}. There it is assumed that Weyl and diffemorphism invariances hold together. Even though full (global) conformal symmetry is known to be in place, in \citep{WeylvsLiouville} it was conjectured that Liouville theory might not be made both diffeomorphic and Weyl invariant, evoking a generic ``classical anomaly'' as the reason for that. In this letter we prove that conjecture and provide explicit formulae for such classical gravitational anomalies. A more detailed discussion on how such classical anomalies arise, in general and for Liouville, can be found in \citep{LargeHamanIorio2023}.

Liouville theory often emerges as an effective action, see, e.g., the recent \citep{Mertens_2021}. Clearly, when a theory is only effective, we can relax symmetry requirements and lack of Weyl or diffeomorphic invariance, or even lack of both, is not a big deal. On the other hand, it is entirely legitimate to study Liouville theory as a fundamental theory, as we do here. In this latter case, the request for a fully symmetric situation, both Weyl and diffeomorphic, is in order.

The concerns of \citep{WeylvsLiouville} are not only important to probe the procedure of \citep{WeylGauging}, but also in the more general investigation of the concept of \textit{anomaly}, in the first place. This is our most important motivation.

As well known, see, e.g., \citep[Chapter~I.2]{DiverseTopics}, it is the process of \textit{second quantization} (fields) rather than first quantization (one particle) that is responsible for the anomaly phenomenon. It must then be seen as a violation occurring while trying to enforce symmetries, valid for certain specific configurations, to setups with infinitely many more allowed configurations.

The question that we pose, and answer, here is whether something along those lines happens when we try to generalize the Lorentz and conformal symmetries of the flat space action \eqref{eq:FlatLiouvilleTheory} to curved space, staying entirely in a classical environment. On top of that, we also find that the classical central charge of the Virasoro algebra plays the same role as in the quantum case. Altogether, this is then clearly a \textit{classical instance} of a phenomenon previously encountered only in the quantum (field theory) context, as conjectured in \citep{WeylvsLiouville}.

As we shall see, all that happens very much according to the original findings of Adler \citep{Adler_1969} and Bell  and Jackiw \citep{BellJackiw_1969} (ABJ) (see also \citep{BetlmannBOOK,DiverseTopics}): \textit{the generalization (to curved space here, to the quantum regime in ABJ) only allows to keep one of the two symmetries valid in the particular regime (flat space here, classical regime in ABJ)}. The nontrivial central charge of the flat theory does not spoil the full (Lorentz and conformal) invariance of the flat space action \eqref{eq:FlatLiouvilleTheory}, but spoils either one or the other symmetry, in curved space. We must have either diffeomorphic (Lorentz \citep{Bardeen_1984}) anomaly or conformal (Weyl) anomaly.

%%%%%%%%%%%%%%%%%%

Let us start by considering the diffeomorphic invariant Liouville action on a curved background
\begin{equation} \label{eq:CurvedLiouville}
  \mathcal A{}_{\scriptscriptstyle{L}}[\Phi]=\int\dd[2]{x}\sqrt{-g}
  \qty(
    \frac12g^{\mu\nu}\nabla_\mu\Phi\nabla_\nu\Phi
    - \frac{m^2}{\beta^2}e^{\beta\Phi}
    + \frac1\beta R\Phi
  ) \,,
\end{equation}
routinely employed to obtain the energy\hyp{}momentum tensor (EMT)
\begin{equation} \label{eq:CurvedEMTLiouvilleCorrect}
  T^{\mu\nu}_{\scriptscriptstyle{L}} = -\frac{2}{\sqrt{-g}}\fdv{\mathcal{A}_{\scriptscriptstyle{L}}}{g_{\mu\nu}} = \nabla^\mu\Phi\nabla^\nu\Phi -g^{\mu\nu}
  \qty(
    \frac12g^{\alpha\beta}\nabla_\alpha\Phi\nabla_\beta\Phi
    - \frac{m^2}{\beta^2}e^{\beta\Phi} )
    + \frac2\beta\qty(g^{\mu\nu} \nabla_\rho\nabla^\rho - \nabla^\mu\nabla^\nu)\Phi \,,
\end{equation}
that \textit{on-shell} gives
\begin{equation}\label{eq:CurvedTrace}
  {T_{\scriptscriptstyle{L}}}^\mu{}_\mu = \frac2{\beta^2} R \,,
\end{equation}
ensuring a zero trace in the flat limit.

No issue appears to occur in flat space, besides the unpleasant \textit{ad-hoc} nature of this procedure to obtain the improved, i.e. traceless, EMT in flat space $\Theta^{\mu \nu} = \lim_{g_{\mu \nu} \to \eta_{\mu \nu}} T_L^{\mu \nu}$, see later Eq. (\ref{eq:FlatLiouvilleEMTimproved}). Nonetheless, as we shall see in the rest of the paper, nontrivial issues about conformal invariance are indeed present already in flat space, as well as in curved space.

To see that, we first notice that the  \textit{lack of Weyl invariance} of the action (\ref{eq:CurvedLiouville}) and the \textit{ad-hoc} nature of the improvement procedure just recalled, could be both faced at once by employing the approach of \citep{WeylGauging}, based on the Weyl-gauging of the curvilinear expression for the action \eqref{eq:FlatLiouvilleTheory}
\begin{equation} \label{eq:WeylGaugedLiouville}
\mathcal A_{\scriptscriptstyle{W}}[\Phi,W_\mu]
  = \int \dd[2]{x}\sqrt{-g}
  \qty(
    \frac{1}{2}g^{\mu\nu}\nabla_\mu\Phi\nabla_\nu\Phi
    - \frac{m^2}{\beta^2}e^{\beta\Phi}
    +\frac{2}{\beta}\Phi\nabla_\mu W^\mu
    +\frac{2}{\beta^2}g^{\mu\nu}W_\mu W_\nu
    ) \,.
\end{equation}
Since under Weyl transformations, $g_{\mu\nu} \to e^{2\omega}g_{\mu\nu}$ and $\Phi \to \Phi-\frac2\beta\omega$, one has $2\nabla_\mu W^\mu \weyl e^{-2\omega}\qty(2\nabla_\mu W^\mu - 2\nabla_\mu\nabla^\mu \omega)$, this should be compared to $R[g_{\mu\nu}]\weyl e^{-2\omega}\qty(R[g_{\mu\nu}]-2\nabla_\mu\nabla^\mu \omega)$ and the (Ricci gauged, in the language of \citep{WeylGauging}) action
\begin{equation} \label{eq:RicciGaugedLiouville}
  \mathcal A_{\scriptscriptstyle{R}}[\Phi] = \int \dd[2]{x}\sqrt{-g}
  \qty(
    \frac{1}{2}g^{\mu\nu}\nabla_\mu\Phi\nabla_\nu\Phi
    - \frac{m^2}{\beta^2}e^{\beta\Phi}
    +\frac{1}{\beta}\Phi R
    +\frac{2}{\beta^2}g^{\mu\nu}W_\mu W_\nu) \,,
\end{equation}
enjoys Weyl invariance, ${T_{\scriptscriptstyle{R}}}^\mu{}_\mu = 0$, provided
\begin{equation} \label{eq:WeylRicci2DIdentification}
2\nabla_\mu W^\mu = R \,,
\end{equation}
holds. Notice that, contrary to \citep{WeylGauging}, we keep here the last, $\Phi$-independent term, that is precisely the one that ensures Weyl invariance.

A solution of \eqref{eq:WeylRicci2DIdentification} can be found \citep{WeylvsLiouville} using the Green's function $K(x,y)$, such that
$\nabla^2_x K(x,y) = \frac{1}{\sqrt{-g(x)}}\delta^{(2)}(x-y)$. Assuming that $W_\mu = \partial_\mu w$, with $w$ transforming as $w \weyl w -\omega$, the solution is
$ w(x) = 1/2 \int\dd[2]{y}\sqrt{-g(y)} \, K(x,y) \, R(y)$. It follows that the extra term in the action proportional to $W^\mu W_\mu$ is
\begin{equation} \label{eq:PolyakovEffectiveAction}
  \int \dd[2]{x} W^\mu(x) W_\mu(x)
  =
  \frac14\int \dd[2]{x}\dd[2]{y} \sqrt{-g(x)}R(x)K(x,y)\sqrt{-g(y)}R(y) \,,
\end{equation}
which is the well-known \textit{Polyakov string effective action} \citep{Polyakov_1981}. The EMT associated to the action (\ref{eq:RicciGaugedLiouville}) with (\ref{eq:PolyakovEffectiveAction}), is traceless and covariantly conserved\footnote{In the quantum case, there is no regularization
for which Weyl and diffeomorphism invariance hold together \cite{LEUTWYLER}.} \citep{Polyakov_1981}. The price we pay is the evident nonlocality.

A \textit{local solution} to (\ref{eq:WeylRicci2DIdentification}) was found by Deser and Jackiw\footnote{Deser and Jackiw arrived at this local solution precisely by looking into the Polyakov action and its associated EMT. A particular solution for Reissner-Nordström black hole was derived by Iso et al in \citep{Iso2006PRD} using the Green's function approach.} in \citep{DeserJackiw96}
\begin{equation} \label{eq:JackiwSolution}
    W^\mu_{\scriptscriptstyle{DJ}} =
    \frac{\varepsilon^{\mu\nu}}{2\sqrt{-g}}\qty[
    \frac{\varepsilon^{\alpha\beta}}{\sqrt{-g}}
    % \partial_\alpha g_{\beta\nu}
    \Gamma_{\beta\alpha\nu} % alternativni formulace
    +
    % \frac{\varepsilon^{\mu\nu}}{2\sqrt{-g}}
    (\cosh\sigma -1)\partial_\nu\gamma + \partial_\nu r]\,,
\end{equation}
where $\varepsilon^{01}=+1$ is the Levi-Civita symbol\footnote{Additionally, we choose the light-cone coordinates such that $\varepsilon^{-+}=+1$, as explained in Appendix \ref{sec:LightConeCoordinates}.} and a ``conformal'' parametrization of the metric gives $g_{++}/\sqrt{-g} =  e^\gamma\sinh\sigma$, $g_{+-}/\sqrt{-g} = \cosh\sigma$, $g_{--}/\sqrt{-g} = e^{-\gamma}\sinh\sigma$, and, from there, $\gamma = \ln\sqrt{g_{++}/{g_{--}}}$ (see Appendix \ref{sec:ImprovementTrace}). The expression \eqref{eq:JackiwSolution} includes the derivative of a generic Weyl scalar, $r$, to take into account the invariance of \eqref{eq:WeylRicci2DIdentification} for $W^\mu_{\scriptscriptstyle{DJ}} \goto W^\mu_{\scriptscriptstyle{DJ}} + \frac{\varepsilon^{\mu\nu}}{2\sqrt{-g}}\partial_\nu r$.

$W^\mu_{\scriptscriptstyle{DJ}}$ enjoys proper Weyl transformations,
$
  g_{\mu\nu}W_{\scriptscriptstyle{DJ}}^\nu \weyl g_{\mu\nu}W_{\scriptscriptstyle{DJ}}^\nu - \partial_\mu\omega \,,
$
but it does not transform as a general (contravariant) vector under infinitesimal diffeomorphisms, $x^\mu \goto x'^\mu = x^\mu - f^\mu(x)$,
\begin{equation} \label{eq:JackiwSolutionDiffeoTransf}
  W'{}_{\scriptscriptstyle{DJ}}^\mu(x') = ~\pdv{x'^\mu}{x^\nu}W_{\scriptscriptstyle{DJ}}^\nu(x)
  + \frac{\varepsilon^{\mu\nu}}{2\sqrt{-g}}\partial_\nu
  \qty[
    \qty(
      \partial_--e^{-\gamma}\tanh \frac\sigma2 \, \partial_+
    )f^-
    -
    \qty(
      \partial_+-e^{\gamma}\tanh \frac\sigma2 \, \partial_-
    )f^+
  ] \,.
\end{equation}
It follows % from \eqref{eq:JackiwSolutionDiffeoTransf}
that the term $g_{\mu\nu}W_{\scriptscriptstyle{DJ}}^\mu W_{\scriptscriptstyle{DJ}}^\nu$ in (\ref{eq:RicciGaugedLiouville}), although it keeps Weyl invariance and locality of
$\mathcal A_{\scriptscriptstyle{R}}[\Phi]$, cannot be a world scalar, hence it breaks diffeomorphism invariance.

To investigate and quantify such breaking, let us start with the contribution to the EMT coming from the extra term
\begin{equation}
  T^{\mu\nu}_{\text{extra}}\equiv
  -\frac{2}{\sqrt{-g}}\fdv{g_{\mu\nu}} \int \dd[2]{x}\frac{2}{\beta^2}\sqrt{-g}W^\mu W_\mu \,.
\end{equation}
In Appendix \ref{sec:ImprovementTrace} it is shown that
\begin{equation}
    \frac{\beta^2}{2}T^{\mu\nu}_{\text{extra}}
    =
      g^{\mu\nu}W_\rho W^\rho
    - 2 W^\mu W^\nu
    - Rg^{\mu\nu}
    + \nabla^\mu W^\nu
    + \nabla^\nu W^\mu
    + 2\frac{\varepsilon^{\alpha\beta}}{\sqrt{-g}}
    % \partial_\beta \qty(g_{\alpha\lambda}W^\lambda)
    \partial_\beta W_\alpha
    [(\cosh\sigma-1)\Gamma^{\mu\nu}
    + r^{\mu\nu} % ambiguity part
    ]\,,
\end{equation}
where $W_\rho\equiv g_{\rho\lambda}W^\lambda$,
$
 2 \Gamma^{\mu\nu}
  =
  (g_{--}g_{++})^{-1/2}
  \begin{pmatrix}
    -\sinh \gamma & \cosh\gamma \\
    \cosh\gamma & -\sinh\gamma
  \end{pmatrix}
$ and $r^{\mu\nu}\equiv \delta r / \delta g_{\mu\nu}$. One would like to compute $\nabla{}_\mu  T{}^{\mu\nu}_{\text{extra}}$ and compare it with known expressions of the \textit{quantum} gravitational anomalies, such as
the so-called \textit{consistent} anomaly \cite{BetlmannBOOK} (our notation follows \citep{Wilczek_2005}, see also \citep{BanerjeeKulkarni})
$\nabla_\mu T^\mu{}_\nu = \frac{1}{48\pi}\frac{\varepsilon^{\sigma\rho}}{2\sqrt{-g}}\partial_\rho\partial_\lambda\Gamma^\lambda{}_{\nu\sigma}$ or the so-called \textit{covariant} anomaly \cite{BetlmannBOOK} (our notation follows \citep{Jackiw_1995}) $\nabla_\mu T^\mu{}_\nu = \frac{1}{48\pi}\partial_\nu R$.

Rather than attempting a direct computation, we take a simpler road. First, we move to isothermal light-cone coordinates, that we know to always exist in two dimensions. There one has
$
  \hat g_{\pm\pm}(x) = e^{2\rho(x)}
  \begin{pmatrix}
    0 & 1 \\
    1 & 0
  \end{pmatrix}
$, and we indicate with a hat all quantities evaluated there\footnote{For this choice, $\varepsilon^{\alpha\beta}\partial_\beta\hat W{}_\alpha = 0$.}. If we set $r=0$ for a moment, we have
\begin{equation}
  \frac{\beta^2}{4}\hat T{}^{\mu\nu}_{\text{extra}}
  =
  (\hat g^{\mu\alpha}\partial_\alpha\rho)(\hat g^{\nu\beta}\partial_\beta\rho)
  -\frac12\hat g^{\mu\nu}(\hat g^{\alpha\beta}\partial_\alpha\rho\partial_\beta\rho)
  +\qty[\hat g^{\mu\nu}(\hat g^{\alpha\beta}\partial_\alpha\partial_\beta)-\hat g^{\mu\alpha}\hat g^{\nu\beta}\partial_\alpha\partial_\beta]\rho\,,
\end{equation}
and the computation becomes trivial, $\hat\nabla{}_\mu \hat T{}^{\mu\nu}_{\text{extra}} = 0$. Of course, this would not guarantee general covariance, until we have a frame-independent result (see later).

On the other hand, including $r$ in the computation gives
\begin{equation} \label{eq:EMTIsothermalDivergence}
    \beta^2\hat\nabla{}_\mu \hat T{}^{\mu\nu}_{\text{extra}} =
    % 0. % r = 0 result
    \frac{\varepsilon^{\nu\mu}}{\sqrt{-\hat g}}\hat g^{\alpha\beta}\partial_\alpha\partial_\beta\partial_\mu \hat r
    -2 \hat g^{\alpha\beta}\partial_\mu\qty(\hat r^{\mu\nu}\partial_\alpha\partial_\beta \hat r)
    + 2\hat r^{\mu\nu}\partial_\mu \hat g^{\alpha\beta}\partial_\alpha\partial_\beta \hat r\,,
\end{equation}
and this  expression, although it differs from the recalled anomalous quantum expressions \citep{Wilczek_2005,Jackiw_1995}, it is clearly nonzero, in general. In this coordinate frame, $\hat T{}^{\mu\nu}_{\text{extra}}$ not only guarantees Weyl invariance, through a traceless EMT, but for harmonic %\footnote{Since harmonic $r$s may lead to Weyl and diffeomorphism invariances to hold together, in the rest of this letter we shall assume that choice.}
$r$s, $\hat\Box \hat r = 0$
\begin{equation} \label{eq:EMTIsothermalDivergenceRVanish}
  \hat\nabla{}_\mu \hat T{}^{\mu\nu}_{\text{extra}}\eval_{\hat\Box \hat r=0} = 0 \,.
\end{equation}
As for the previous case, for $r=0$, this is not enough to have general covariance. We have no guarantee that \eqref{eq:EMTIsothermalDivergenceRVanish} holds in all frames. We have to look for how much such divergence differs from a tensor, when we move away from the isothermal frame,
$
  \Delta \hat\nabla{}_\mu \hat T{}^{\mu\nu}_{\text{extra}}(x)
  \equiv
  \nabla'_\mu T'{}^{\mu\nu}_{\text{extra}}(x')
  - (\partial x'^\nu / \partial x^\rho) \hat\nabla{}_\sigma \hat T{}^{\sigma\rho}_{\text{extra}}(x)$. This has to be, at least partially, expressible in terms of
$\Delta \hat W{}^{\mu}(x)
  \equiv
  W'^{\mu}(x')
  - (\partial x'^{\mu} / \partial x^\nu ) \hat W{}^\nu(x)
$. For infinitesimal diffeomorphisms, $W^\mu$ transforms as \eqref{eq:JackiwSolutionDiffeoTransf} and, defining $\Delta r(x) \equiv r'(x') - r(x)$,
\begin{equation}\label{eq:NonTensorialWTransf}
    \Delta \hat W^\mu
    =
    % \frac{\varepsilon^{\mu\nu}}{2\sqrt{-\hat g}}\partial_\nu
    % \qty(
    %   \partial_-f^- - \partial_+f^+ + \Delta\hat r
    % )
    % =
    \frac{\varepsilon^{\mu\nu}}{2\sqrt{-\hat g}}\partial_\nu
    \qty[
      % \partial_\alpha\qty(\frac{\varepsilon^{\alpha\beta}}{\sqrt{-\hat g}}f_\beta)
      \qty(\partial_-f^--\partial_+f^+)
      +\Delta\hat r
    ]
    \equiv
    \frac{\varepsilon^{\mu\nu}}{2\sqrt{-\hat g}}\partial_\nu \xi(r,f)\,.
\end{equation}
With these
\begin{equation}\label{eq:DivergenceEMTtransformation}
    \beta^2\Delta\hat\nabla{}_\mu \hat T{}^{\mu\nu}_{\text{extra}}
    =
    \frac{\varepsilon^{\nu\mu}}{\sqrt{-\hat g}}
    \hat g^{\alpha\beta}
    \partial_\alpha\partial_\beta\partial_\mu\xi(r,f)
    % \varepsilon^{\sigma\rho}
    % \partial_\sigma\qty(\frac{f_\rho}{\sqrt{-\hat g}})
    -2\hat g^{\alpha\beta}\partial_\mu\qty(\hat r^{\mu\nu} \partial_\alpha\partial_\beta\xi(r,f))
    +2\hat r^{\mu\nu}\partial_\mu \hat g{}^{\alpha\beta}\partial_\alpha\partial_\beta\xi(r,f)\,,
\end{equation}
that, for the choice \eqref{eq:JackiwSolution}, and for $r=0$, eventually gives a compact expression
\begin{equation}\label{eq:rzerochoice}
    \Delta\hat\nabla{}_\mu \hat T{}^{\mu\nu}_{\text{extra}}
    = \frac{1}{\beta^2} \,
    \frac{\varepsilon^{\nu\mu}}{\sqrt{-\hat g}}
    \hat g^{\alpha\beta}
    \partial_\alpha\partial_\beta
    \partial_\mu
    \qty(\partial_-f^--\partial_+f^+)
    \,.
\end{equation}
This expression does not vanish for a general $f^\mu$. \textit{This proves the loss of diffeomorphism invariance in the Weyl invariant formulation of Liouville theory \eqref{eq:RicciGaugedLiouville}, with local solution \eqref{eq:JackiwSolution}}.

Quadratic transformations, $f^\mu = a^{\mu}{}_{\alpha\beta}x^\alpha x^\beta + b^\mu{}_\alpha x^\alpha + c^\mu$, which include Poincaré transformations, $f^\mu = \omega^\mu{}_\nu x^\nu + c^\mu$, preserve the tensorial nature of $\hat\nabla_\mu \hat T{}^{\mu\nu}_{\text{extra}}$ and so do conformal transformations, obeying $\Box f^\mu = 0$. Therefore, for infinitesimal conformal and Poincaré transformations in flat space, the extra term, $T^{\mu\nu}_{\text{extra}}$, is covariantly conserved, regardless of the choice of $r$. In other words, $T^{\mu\nu}_{\text{extra}}$ does not violate the symmetries of $T^{\mu\nu}$ in the flat limit, that is the same conclusion of \citep{WeylGauging}.

To complete the proof that this lack of diffeomorphic invariance is indeed the classical version of the quantum anomaly, we need to relate it to a nontrivial center of the Virasoro algebra. To do so, let us first consider the flat limit of the EMT (\ref{eq:CurvedEMTLiouvilleCorrect})
\begin{equation}\label{eq:FlatLiouvilleEMTimproved}
  \Theta_{\mu\nu} = \partial_\mu\Phi\partial_\nu\Phi -\eta_{\mu\nu}\qty(
    \frac{1}{2}\partial_\alpha\Phi\partial^\alpha\Phi - \frac{m^2}{\beta^2}e^{\beta\Phi})
  + \frac{2}{\beta}(\eta_{\mu\nu}\Box - \partial_\mu\partial_\nu)\Phi \,,
\end{equation}
that is traceless on-shell. The associated Noether charges, written in the light-cone frame\footnote{Light-cone coordinates are defined in the Appendix \ref{sec:LightConeCoordinates}. Further details on the expression of these charges in this frame can be found in \citep{LargeHamanIorio2023}}
\begin{equation}\label{eq:LiouvilleNoetherCharge}
   Q^\pm[f] = \int \dd{x^\pm} \Theta_{\pm\pm}f^\pm =\int \dd{x^\pm} \qty((\partial_\pm\Phi)^2-\frac2\beta\partial^2_\pm\Phi)f^\pm \,,
\end{equation}
through the Poisson brackets $\eval{\pb{\Phi(x)}{\Phi(y)}}_{x^+=y^+} = - \frac14\sgn(x^--y^-)$ and $\eval{\pb{\Phi(x)}{\Phi(y)}}_{x^-=y^-} =  -\frac14\sgn(x^+-y^+)$, generate the right transformations
\begin{equation}
  \delta_f  \Phi \equiv \pb{\Phi(x^+,x^-)}{ Q^\pm[f]}
  =  f^\pm(x^\pm)\partial_\pm \Phi(x^+,x^-)
  + \frac1\beta \partial_\pm f^\pm(x^\pm)\,.
\end{equation}
They are made of two terms, both necessary for the invariance of the flat action (\ref{eq:FlatLiouvilleTheory}): the usual Lie derivative of the scalar field, $f^\alpha \partial_\alpha \Phi$, and an affine term.
It is easy to verify that these charges obey an algebra with a genuine central extension
\begin{equation}\label{eq:LiouvilleChargeAlgebra}
  \pb{ Q^\pm[f]}{ Q^\pm[g]} =
  Q^\pm[k] + \frac{1}{\beta^2}\Delta^\pm[f,g] \,,
\end{equation}
where $k^\mu = f^\nu\partial_\nu g^\mu- g^\nu\partial_\nu f^\mu$ and $\Delta^\pm[f,g] = \int \dd{x^\pm} (g^\pm\partial_\pm^3f^\pm- f^\pm\partial_\pm^3g^\pm)$. By restricting to a periodic manifold, with a periodicity $P$, $x^\pm \propto x^\pm + P$, generators can be decomposed into
\begin{equation}\label{eq:LiouvilleGenerator}
   Q^\pm_n \equiv \frac{P}{2\pi}\int \dd{x^\pm} \Theta_{\pm\pm} e^{i\frac{2\pi}{P}nx^\pm}= \frac{P}{2\pi} Q^\pm[e^{i\frac{2\pi}{P}nx^\pm}] \,,
\end{equation}
and the algebra \eqref{eq:LiouvilleChargeAlgebra} can be recast into the following form
\begin{equation}\label{eq:LiouvilleVirasoroAlgebra}
  i\pb{ Q^\pm_n}{ Q^\pm_m} = (n-m) Q^\pm_{n+m} + \frac{4\pi}{\beta^2}n^3\delta_{n+m,0} \,
\end{equation}
that is just the Virasoro algebra with genuine central charge
\begin{equation}\label{eq:Center}
  c = \frac{48 \pi}{\beta^2} \,.
\end{equation}
It is the algebra of the flat Liouville EMT components that inevitably includes the genuine center \eqref{eq:Center}
\begin{equation}
  \pb{\Theta_{\pm\pm}(x)}{\Theta_{\pm\pm}(y)}\eval_{x^\mp=y^\mp}
  =
  \Theta'_{\pm\pm}(x)\delta(x^\pm-y^\pm)
  +2\Theta_{\pm\pm}(x)\delta'(x^\pm-y^\pm)
  -\frac c{24\pi}\delta'''(x^\pm-y^\pm) \,,
\end{equation}
and so do its transformations
\begin{equation}\label{eq:InfinitesimalEMTtransformation}
  \delta_f  \Theta_{\pm\pm}
  = f^\pm\partial_\pm \Theta_{\pm\pm}
  +2 \Theta_{\pm\pm}\partial_\pm f^\pm
  % -\frac2{\beta^2}\partial^3_\pm f^\pm
  -\frac c{24\pi}\partial^3_\pm f^\pm \,.
\end{equation}
This center is not there in the trace of flat EMT (\ref{eq:FlatLiouvilleEMTimproved}), but it is proportional to the trace \eqref{eq:CurvedTrace} of the curved space EMT \eqref{eq:CurvedEMTLiouvilleCorrect}.

A deeper study of this, including the general framework for classically anomalous transformations, we have done it in \cite{LargeHamanIorio2023}. Here we want to show that the above is indeed related to the extra term $T^{\mu\nu}_{\text{extra}}$ that, in curved space, preserves Weyl invariance but breaks diffeomorphic invariance. To do so, let us first rewrite \eqref{eq:InfinitesimalEMTtransformation} as the difference between the full transformation and its tensorial part
\begin{equation}\label{eq:NonTensorialEMTtransformation}
  \Delta \Theta_{\pm\pm} \equiv \delta_f  \Theta_{\pm\pm}
  - f^\pm\partial_\pm \Theta_{\pm\pm}
  - 2 \Theta_{\pm\pm}\partial_\pm f^\pm
  =
  -\frac c{24\pi}\partial^3_\pm f^\pm \,,
\end{equation}
as we did earlier in the curved context. We then simply notice that, for the infinitesimal diffeomorphism, $x^\mu\goto x^\mu - f^\mu(x)$, the non-tensorial transformation of $T^{\mu\nu}_{\text{extra}}$ is
\begin{equation}
  \beta^2\Delta\hat T{}^{\mu\nu}_{\text{extra}}(x)
  =\partial_\alpha \partial_\beta \xi(r,f)
  \qty(
    \hat g^{\mu\alpha}\frac{\varepsilon^{\nu\beta}}{\sqrt{-\hat g}}
    + \hat g^{\nu\alpha}\frac{\varepsilon^{\mu\beta}}{\sqrt{-\hat g}}
  )
  % +2\hat g^{\mu\nu}\frac{\varepsilon^{\alpha\beta}}{\sqrt{-\hat g}}\partial_\alpha \rho\partial_\beta\xi(r,f)
  \,,
\end{equation}
where the same notation of \eqref{eq:NonTensorialWTransf} and \eqref{eq:DivergenceEMTtransformation} has been used. Assuming conformal diffeomorphisms and taking the flat limit we have
\begin{equation} \label{eq:DiffeoAnomaly}
\Delta\hat T{}^{\pm\pm}_{\text{extra}}(x)\eval_{\rho\goto0}
= - \frac2{\beta^2} \, \partial^3_\mp f^\mp = \Delta \Theta_{\mp\mp}\,,
\end{equation}
which is exactly\footnote{Due to the signature of the light-cone metric, $\Theta_{\mp\mp} = \Theta^{\pm\pm}$. See Appendix \ref{sec:LightConeCoordinates}.} \eqref{eq:NonTensorialEMTtransformation} with $c$ given by \eqref{eq:Center}.

This center was removed from the trace \eqref{eq:CurvedTrace} but re-emerged in \eqref{eq:DiffeoAnomaly}. We have then proved that, also for \textit{classical} Liouville theory, lack of Weyl invariance or of diffeomorphism invariance is related to the Virasoro center, like in the quantum case. This gives a precise mathematical meaning to what we are now entitled to call ``classical gravitational anomalies''. Whether this is possible for more general classical systems, it is an important open question. Another direction for further research that we are considering is the connection of such anomalous transformation of the EMT with ``classical Unruh and Hawking effects''.

\textit{Acknowledgments.} We gladly acknowledge support from Charles University Research Center (UNCE/SCI/013).

\bibliographystyle{apsrev4-2}
% \bibliography{../bibliography_Liouville}


\newpage

\widetext
\begin{center}
  \textbf{\Large Appendices}
\end{center}
\appendix

\setcounter{equation}{0}
\setcounter{figure}{0}
\setcounter{table}{0}

\section{Light-cone coordinates}\label{sec:LightConeCoordinates}
We define  the light-cone coordinates in two dimensions as
\begin{equation}
  x^\pm=\frac{1}{\sqrt{2}} \big(x^0\pm x^1\big) \,.
\end{equation}

The derivatives in ligh-cone coordinates are
\begin{equation}
    \partial_+ = \frac{1}{\sqrt{2}}(\partial_0 + \partial_1)
    \qc
    \partial_- = \frac{1}{\sqrt{2}}(\partial_0 - \partial_1) \,.
\end{equation}

The coordinate transformation $(x^0,x^1)\goto(x^-,x^+)$, where we set the index ``-'' to be the first one, can be obtained by acting with the matrix
\begin{equation}
  S^\mu{}_\nu \equiv \pdv{x'^\mu}{x^\nu} = \frac{1}{\sqrt{2}}
    \left(
      \begin{matrix}
        1 & -1 \\
        1 & 1
      \end{matrix}
    \right) \,.
\end{equation}
This choice gives a unit determinant, $\det S^\mu{}_\nu = +1$, and this transformation does not add any additional factors to transformations of tensor densities, e.g. $\sqrt{-g}$ and $\varepsilon^{\mu\nu}$.

Light-cone components of the metric are, thus, obtained as
\begin{align}
  g_{++}&=\frac{1}{2} \big( g_{00}+2g_{01}+g_{11} \big)
  \\
  g_{+-}&=\frac{1}{2} \big( g_{00}-g_{11} \big)
  \\
  g_{--}&=\frac{1}{2} \big( g_{00}-2g_{01}+g_{11} \big) \,,
\end{align}
and the components of inverse metric $g^{\mu\nu}$ are
\begin{equation}
    g^{++}=\frac{g_{--}}{g}\qc
    g^{+-}=-\frac{g_{+-}}{g}\qc
    g^{--}=\frac{g_{++}}{g} \,.
\end{equation}

The light-cone Minkowski metric, with the signature $\eta_{\mu \nu} = {\rm diag} (+1,-1)$, is
\begin{equation}
  \eta_{\mu\nu} = \left(
  \begin{array}{cc}
    0 & 1  \\
    1 & 0
  \end{array}
  \right) \,,
\end{equation}
with $\mu,\nu \in \{-,+\}$. Thus, the scalar product is $x^2 = 2x^+x^-$. Raising and lowering indices change the $+$ to $-$ sign and vice versa, e.g. $\partial_\pm = \partial^\mp$, $\Theta_{\mp\mp} = \Theta^{\pm\pm}$. The Levi-Civita symbol is $\varepsilon^{-+}=+1$.

\section{Improvement of the Energy-momentum Tensor}\label{sec:ImprovementTrace}
Here we compute the extra improvement term of the EMT
\begin{equation*}
	T^{\mu\nu}_{\text{extra}}\equiv
	-\frac{2}{\sqrt{-g}}\fdv{g_{\mu\nu}} \int \dd[2]{x}\frac{2}{\beta^2}\sqrt{-g}W^\mu W_\mu
  \equiv
  -\frac{2}{\sqrt{-g}}\fdv{\Delta A}{g_{\mu\nu}}
  \,.
\end{equation*}

To facilitate calculations it is easier to introduce the ``conformal'' metric
\begin{equation}
  \gamma_{\mu\nu}\equiv\frac{g_{\mu\nu}}{\sqrt{-g}},
  \qquad
  \gamma^{\mu\nu} \equiv \sqrt{-g} g^{\mu\nu},
  \qquad
  \sqrt{-g} \equiv \rho \,,
\end{equation}
which, in light-cone coordinates, can be parametrized as
\begin{equation}
  \gamma_{++} = e^\gamma \sinh\sigma
  \qc
  \gamma_{+-} = \cosh\sigma
  \qc
  \gamma_{--} = e^{-\gamma} \sinh\sigma \,.
\end{equation}
In this parametrization a new quantity $R^\mu$ can be defined
\begin{equation}
  R^\mu \equiv 2\sqrt{-g}W^\mu \,.
\end{equation}
Using the following identity
\begin{equation}
  \epsilon^{\mu\nu}\epsilon^{\alpha\beta}
  =
  \gamma^{\mu\beta}\gamma^{\nu\alpha} - \gamma^{\mu\alpha}\gamma^{\nu\beta} \,,
\end{equation}
it can be seen that
\begin{equation}
  R^\mu = -\gamma^{\mu\nu}\partial_\nu \rho - \partial_\nu \gamma^{\mu\nu} + \epsilon^{\mu\nu}(\cosh \sigma-1)\partial_\nu\gamma + \epsilon^{\mu\nu}\partial_\nu r \,.
\end{equation}

The natural way to compute the EMT for $\Delta A$ is by varying the latter with respect to $\gamma^{\mu\nu}$
\begin{equation}
  T_{\mu\nu} = \frac{2}{\sqrt{-g}}\frac{\delta \Delta A}{\delta g^{\mu\nu}}
  = 2\frac{\delta \Delta A}{\delta \gamma^{\mu\nu}}
  - \gamma_{\mu\nu}\gamma^{\alpha\beta}\frac{\delta \Delta A}{\delta \gamma^{\alpha\beta}} \,.
\end{equation}
To begin, we see that
\begin{equation}
  \delta \Delta A =
  \frac{1}{2\beta^2}\integral{2} \delta \gamma_{\mu\nu}R^\mu R^\nu
  + \frac{1}{\beta^2}\integral{2} \gamma_{\mu\nu}R^\nu \delta R^\mu \,,
\end{equation}
where
\begin{equation}
  \delta R^\mu = -\delta \gamma^{\mu\nu}\partial_\nu \sigma
  - \gamma^{\mu\nu}\partial_\nu\delta\sigma - \partial_\nu\delta\gamma^{\mu\nu}
  + \partial_\nu[\epsilon^{\mu\nu}(\cosh\sigma-1)\delta\gamma]
  - \bar{\Gamma}^\mu_{\alpha\beta}\delta\gamma^{\alpha\beta} + \epsilon^{\mu\nu}\partial_\nu\delta r \,.
\end{equation}
The last two terms are the result of the following computation \citep{DeserJackiw96}
\begin{equation}
    \delta\qty[\epsilon^{\mu\nu}(\cosh\sigma - 1)\partial_\nu\gamma]
    - \partial_\nu
    \qty[\epsilon^{\mu\nu}(\cosh\sigma - 1)\delta\gamma]
    =- \bar{\Gamma}^\mu_{\alpha\beta}\delta\gamma^{\alpha\beta}
    \delta\gamma^{\alpha\beta} \,,
\end{equation}
where
\begin{equation}
  \bar{\Gamma}^\mu_{\alpha\beta} = \frac{1}{2}\gamma^{\mu\nu}
  \left(
    \partial_\alpha\gamma_{\nu\beta} + \partial_\beta\gamma_{\nu\alpha} - \partial_\nu\gamma_{\alpha\beta}
  \right) \,.
\end{equation}
Let us now split $\delta\Delta A$ into four terms
\begin{equation}
  \delta\Delta A = \delta\Delta A^1 + \delta\Delta A^2 + \delta\Delta A^3
  +\delta\Delta A^4 \,,
\end{equation}
where
\begin{equation*}
  \begin{aligned}
    \delta \Delta A^1 =&  \frac{1}{2\beta^2} \integral{2}
    \delta\gamma_{\mu\nu}R^\nu R^\mu \,,
    \\
    \delta \Delta A^2 =&  \frac{1}{\beta^2} \integral{2}
    \gamma_{\mu\nu}R^\nu
    (
      - \delta\gamma^{\mu\lambda}\partial_\lambda\rho
      - \partial_\lambda\delta\gamma^{\mu\lambda}
      - \bar{\Gamma}^\mu_{\alpha\beta}\delta\gamma^{\alpha\beta}
    )
    \\
    =& \frac{1}{2\beta^2} \integral{2}
    \delta\gamma^{\alpha\beta}
    \left[
      g_{\beta\lambda}\nabla_\alpha \left(\frac{R^\lambda}{\sqrt{-g}}\right)
      + g_{\alpha\lambda}\nabla_\beta \left(\frac{R^\lambda}{\sqrt{-g}}\right)
    \right] \,,
    \\
    \delta \Delta A^3 =&  -\frac{1}{\beta^2} \integral{2}
    R^\mu\partial_\mu\delta \rho = -\frac{1}{2\beta^2} \integral{2}
    \sqrt{-g} R g_{\alpha\beta}\delta g^{\alpha\beta} \,,
    \\
    \delta\Delta A^4 =& \frac{1}{\beta^2} \integral{2}
    R^\mu\gamma_{\mu\nu}
    \{
      \partial_\lambda[\epsilon^{\nu\lambda}(\cosh\omega-1)\delta\gamma]
      + \epsilon^{\nu\lambda}\partial_\lambda\delta r
    \}
    \\
    =& -\frac{1}{\beta^2} \integral{2}
    \partial_\lambda (R^\mu\gamma_{\mu\nu})\epsilon^{\nu\lambda}\delta g_{\alpha\beta}
    [(\cosh\omega-1)\Gamma^{\alpha\beta} + r^{\alpha\beta}] \,.
  \end{aligned}
\end{equation*}
TO derive the second expression for $\delta\Delta A^2$ we used
\begin{equation*}
  \begin{split}
    \partial_\mu T^{\alpha\ldots}_{\beta\ldots} + \partial_\mu\rho T^{\alpha\ldots}_{\beta\ldots}
    =&~ \frac{1}{\sqrt{-g}}\partial_\mu (\sqrt{-g} T^{\alpha\ldots}_{\beta\ldots}) \,,
    \\
    g_{\alpha\lambda}\nabla_\beta V^\lambda + g_{\beta\lambda}\nabla_\alpha V^\lambda
    =&~ g_{\alpha\lambda}\partial_\beta V^\lambda + g_{\beta\lambda}\partial_\alpha V^\lambda
    + V^\lambda \partial_\lambda g_{\alpha\beta} \,,
    \\
    \gamma_{\alpha\beta}\delta\gamma^{\alpha\beta} =&~ 0 \,.
  \end{split}
\end{equation*}
To derive $\delta\Delta A^3$ we used the fact that $R^\mu$ is a solution of
\begin{equation*}
  \sqrt{-g} R = \partial_\mu R^\mu \,.
\end{equation*}
Finally, rewriting $\delta\Delta A^4$, we defined
\begin{equation*}
  \begin{aligned}
    \delta\gamma =&~ \Gamma^{\mu\nu}\delta g_{\mu\nu} \,,
    \\
    \delta r =&~ r^{\mu\nu}\delta g_{\mu\nu} \,.
  \end{aligned}
\end{equation*}
By a direct calculations it follows that
\begin{equation}
  \Gamma^{\mu\nu} = \frac{1}{2\sqrt{g_{--}g_{++}}}
  \left(
  \begin{matrix}
    -\sinh \gamma & \cosh\gamma \\
    \cosh\gamma & -\sinh\gamma
  \end{matrix}
    \right) \,.
\end{equation}

With this preliminaries, we are now ready for the computation of the EMT in four steps
\begin{equation*}
  T^i_{\mu\nu} = \frac{2}{\sqrt{-g}}\frac{\delta \Delta A^i}{\delta g^{\mu\nu}} \,,
\end{equation*}
with $i=1,2,3,4$, and the results are
\begin{equation}
  \begin{split}
    T^1_{\mu\nu} =&~ \frac{1}{\beta^2}
    \left(
      \frac{1}{2}\gamma_{\mu\nu}\gamma_{\alpha\beta}R^\alpha R^\beta
      -\gamma_{\mu\alpha}\gamma_{\nu\beta}R^\alpha R^\beta
    \right) \,,
    \\
    T^2_{\mu\nu} =&~ \frac{1}{\beta^2}
    \left(
      g_{\mu\lambda}\nabla_\nu \left(\frac{R^\lambda}{\sqrt{-g}}\right)
      + g_{\nu\lambda}\nabla_\mu \left(\frac{R^\lambda}{\sqrt{-g}}\right)
    \right)
    -\frac{1}{\beta^2}Rg_{\mu\nu} \,,
    \\
    T^3_{\mu\nu} =&~ -\frac{1}{\beta^2}Rg_{\mu\nu} \,,
    \\
    T^4_{\mu\nu} =&~ \frac{2}{\beta^2\sqrt{-g}}
    \partial_\beta \left(\frac{R^\lambda g_{\alpha\lambda}}{\sqrt{-g}}\right)\epsilon^{\alpha\beta}
    [(\cosh\sigma-1)\Gamma_{\mu\nu} + r_{\mu\nu}] \,.
  \end{split}
\end{equation}

Adding these together we have
\begin{equation}\label{eq:LEMTimprovement}
  \begin{split}
    \beta^2 T^{\mu\nu}_{\text{extra}} =&~ \frac{1}{g}
    \left(
      R^\mu R^\nu - \frac{1}{2} g^{\mu\nu} R \cdot R
    \right)
    - 2Rg^{\mu\nu}
    \\
    &~+ g^{\mu\alpha}\nabla_\alpha\left(\frac{R^\nu}{\sqrt{-g}}\right)
    + g^{\nu\alpha}\nabla_\alpha\left(\frac{R^\mu}{\sqrt{-g}}\right)
    \\
    &~+ \frac{2}{\sqrt{-g}}
    \partial_\beta \left(\frac{R^\lambda g_{\alpha\lambda}}{\sqrt{-g}}\right)\epsilon^{\alpha\beta}
    [(\cosh\sigma-1)\Gamma^{\mu\nu} + r^{\mu\nu}]
    \\
    =&~ 2g^{\mu\nu}W^\alpha W^\beta g_{\alpha\beta} - 4 W^\mu W^\nu - 2Rg^{\mu\nu}
    \\
    &~+ 2g^{\mu\alpha}\nabla_\alpha W^\nu
    + 2g^{\nu\alpha}\nabla_\alpha W^\mu
    \\
    &~+ \frac{4}{\sqrt{-g}}
    \partial_\beta \left(W^\lambda g_{\alpha\lambda}\right)\epsilon^{\alpha\beta}
    [(\cosh\sigma-1)\Gamma^{\mu\nu} + r^{\mu\nu}] \,.
  \end{split}
\end{equation}

Since this improvement term should cancel the trace of the EMT, let us compute its trace
\begin{equation}
  \frac{\beta^2}2 T_{\text{extra}}{}^\mu{}_\mu = -R + 2\frac{\epsilon^{\alpha\beta}}{\sqrt{-g}}
  \partial_\beta \left(W^\lambda g_{\alpha\lambda}\right)
  [(\cosh\sigma-1)\Gamma^{\mu\nu} + r^{\mu\nu}]g_{\mu\nu} \,.
\end{equation}

Recalling that
\begin{equation*}
  \Gamma^{\mu\nu} = \frac{2}{(g_{00}+g_{11})^2-4g_{01}^2}
  \left[
    \frac{1}{2} (\delta_0^\mu\delta_1^\nu + \delta_1^\mu\delta_0^\nu)
    (g_{00}+g_{11})
    - g_{01}(\delta_0^\mu\delta_0^\nu+\delta_1^\nu\delta_1^\mu)
  \right] \,,
\end{equation*}
we see that
\begin{equation}
  g_{\mu\nu}\Gamma^{\mu\nu} = 0 \,,
\end{equation}
therefore
\begin{equation}
  T_{\text{extra}}{}^\mu{}_\mu  = -\frac{2}{\beta^2}R + \frac{4}{\sqrt{-g}}
  \partial_\beta \left(W^\lambda g_{\alpha\lambda}\right)\epsilon^{\alpha\beta}
  r^{\mu\nu}g_{\mu\nu} \,.
\end{equation}
From here we see another condition for $r$
\begin{equation}
  g_{\mu\nu}r^{\mu\nu} = 0 \,,
\end{equation}
with which
\begin{equation}
  T_{\text{extra}}{}^\mu{}_\mu  = -\frac{2}{\beta^2}R \,.
\end{equation}
Hence the trace of the improvement cancels the anomalous trace of the canonical EMT
\begin{equation*}
  T^\mu_\mu = \frac{2}{\beta^2}R \,.
\end{equation*}
This proves the Weyl invariance of the improved Liouville action.

\end{document}